# "On the Correlation of Subevents in the ATLAS and CMS/Totem Experiments"

**by Sebastian White**

Brookhaven National Laboratory
and The Rockefeller University
swhite@BNL.gov
Sebastian.White@Rockefeller.edu

- **Introduction:**

We analyze the problem of correlating pp interaction data from the central detectors with a subevent measured in an independent system of leading proton detectors using FP420 as an example.

FP420 is an R&D project conducted by a collaboration formed by members of ATLAS and CMS to investigate the possibility of detecting new physics in the central exclusive channel,

**PP -> P + X + P,**

where the central system X may be a single particle, for example a Standard Model Higgs boson. With standard LHC optics, the protons emerge from the beam at a distance of 420m from the Interaction Point, for M_X ~ 120 GeV. The mass of the central system can be measured from the outgoing protons alone, with a resolution of order 2 GeV irrespective of the decay products of the central system. In addition, to a very good approximation, only central systems with $0^{++}$ quantum numbers can be produced, meaning that observation of a SM or MSSM Higgs Boson in this channel would lead to a direct determination of the quantum numbers.

The challenging aspect of this search is that the calculation of Khoze et al.[1] yields a cross section for visible Higgs of ~10 fbarn with a factor of 2.5 uncertainty, and recent CDF data, though confirming the calculation for exclusive dijet production, indicates that the cross section corresponds to the low edge of the error band. With an expected integrated Luminosity per year of 30 $\text{fb}^{-1}$ at the maximum design luminosity of $10^{34}\,\text{cm}^{-2}\,s^{-1}$ the yield is small, totalling of order 100 events. This means that FP420 must be designed to correlate the leading protons with interactions observed in the central detector even at 20 or more interactions/crossing.

We first show that this problem is far more difficult than previously thought [2], even with state-of-the-art measurement of time-of-flight of the leading protons. We then introduce the notion of correlating subevents in the time domain in either ATLAS or CMS/Totem Electromagnetic Calorimeters.

This problem addresses the long-standing debate between, for example, Tevatron proponents and Isabelle proponents or more recently those of the former SSC[3]- i.e. whether it is possible to do physics at $L = 10^{34} \mathrm{cm}^{-2} s^{-1}$with, on average, 20 interactions per crossing.

We point out that an underappreciated and unintended by-product of a brilliant analysis of optimal signal filterring in the ATLAS Liquid Argon Calorimeter[4] is likely to be crucial for resolving this problem in ATLAS. ATLAS has shown, using test beam data, that this method, as implemented, has achieved the promised resolution on absolute time of the pp interaction of $\sigma_t$=500 psec/E in the noise term, where E in GeV is the energy deposited by a 60 GeV b quark jet , for example. However, as implemented in the test beam [5] the Liquid Argon timing resolution was limited by a constant term of 70 psec.

CMS/Totem has done very similar studies with their testbeam data [6] and concluded that their ECAL has an ultimate resolution of ~120 psec at this shower energy.

Of course, what the actual time resolution of these detector systems is- or could become with sufficient attention -is a very interesting question. It is entirely possible that the good intrinsic timing resolution of the ECALs has been designed out of the data acquisition system in one or both of the experiments due to lack of attention to this important parameter. We feel that it is too early to arrive at a conclusive answer. So we consider three cases for this parameter- an ideal case equal to the noise term of the ATLAS Electromagnetic calorimeter (10 psec), a resolution of 70 psec and a, perhaps more realistic, resolution of 120 psec.

## ■ Method:

To analyze the problem we study the population within an individual beam crossing in both the spatial domain (z-vertex) and the time domain (t-absolute). We first remind ourselves that, in the traversal of two gaussian beam distributions, the spatial distribution of the Luminosity function, L(z), is an invariant with respect to time. Therefore the measurements by the central detector, of the interaction distribution within a crossing are uncorrelated in space and time.

ie
$$L(z,t)=I(z,t)*I(z,-t)=\frac{e^{-\frac{(-c\,t+z)^2}{2\sigma_1^2}-\frac{(c\,t+z)^2}{2\sigma_1^2}}}{2\pi\sigma_1^2}=\frac{e^{-\frac{c^2 t^2+z^2}{\sigma_1^2}}}{2\pi\sigma_1^2}=L(z)*L(t)$$

which manifestly factors into the product of 2 Gaussian distributions- one in space with a variance of $\frac{\sigma_1}{\sqrt{2}}$ and one in time with a variance of $\frac{\sigma_1}{c\sqrt{2}}$.

To calculate the probability distribution of interactions within a crossing we consider the LHC beam and collision parameters at IP1,5 [7]. The distribution is governed by the time and space distribution of the individual bunch current densities, the betatron amplitude function and the beam crossing angle of 285 microradians. In the central detectors the individual events are characterized by the vertex position $z_{\mathrm{vertex}}$, reconstructed with the inner trackers and by their time, assumed to be reconstructed from the Electromagnetic Calorimeter. In the leading proton detectors these two parameters are reconstructed from the time average and the time difference of the time-of-flight to the 2 proton arms- ie

$$Z_{\mathrm{vertex}} = c * (t_1 - t_2) / 2.$$

Having determined the probability distribution and the Poisson distribution of the number of interactions per crossing, it remains to choose realistic parameters for the experimental resolutions. In the central detector we find $\sigma_{z-\mathrm{vertex}}$=50microns which is, in any case, negligible when compared with the corre-sponding leading proton measurement. For the central detectors we take the above possible parameters of the ECALs to determine the single jet timing resolution. For events with 2 jets, of interest for this problem, we average the two jet time measurements.

The possible time resolution on the leading proton is under study by the FP420 Collaboration but there is evidence in the literature [8] that 10 psec can be achieved. Of course the systems aspects of this measurement are challenging and will have to be addressed. In any case we assume 10 psec.

## Results: Quality factor

We first look at the nearest neighbor distances between events in a crossing in both space and time. Many people find it suprising that the distributions in both Fig.2 and Fig. 4 fall exponentially starting with the highest frequency in the first bin. I thank Sergio Rescia for pointing out that this is a basic result from Queueing Theory [ 9]. For a Poisson Distributed population density the distances fall off exponentially from zero.

We then evaluate a Quality Factor, which is the average probability that, within a crossing, there is a unique association of the leading protons with a central detector event- requiring that the association be the correct one. We evaluate the Quality Factor as a function of the cut applied in the experiment ($1*\sigma$,$2*\sigma$,$3*\sigma$) for each of the 3 cases of ideal, good and modest ECAL time resolution. This quality corresponds roughly to the efficiency with which one can use the available luminosity to extract true signal events for each of the assumptions.

## Results: Contamination Factor:

At the analysis level, the most intractable background for the signal we are looking at is the random overlap of a central collision containing b quark jets, for example, with two leading protons from single diffraction dissociation events [10]. The latter will each have randomly distributed times and consequently a random distribution of the Z-vertex computed from the times (ie Zvertex=c*(t1-t2)/2), which could overlap with the Central detector event. In this case we do not have to consider all events in the crossing but only one of them in which the b quark jets were found.

## Results: 10 vs 20 psec in the leading proton time of flight:

We then illustrate the importance of achieving the best possible time of flight measurement on the leading protons. We repeat the contamination factor calculation when the proton resolution is 20 psec rather than 10 psec. We use the case of good but not ideal ECAL resolution.

## Conclusion:

In the search for exclusive Higgs production both the Quality Factor and the Contamination Factor are of interest. Roughly speaking the figure of merit of the experiment is given by Figure-of-Merit= Quality Factor/Contamination Factor and therefore even a factor of 2 gain in either of these due to improved detector resolution will correspond to a factor of 4 in the Figure of Merit. We are presently doing a detailed study of the ATLAS Liquid Argon timing performance. This will be discussed in a future paper.

### Source Code:

The source code for this mathematica notebook is available at http://b5.rockefeller.edu/code/sebastianwhite/ . It can be run using the free *Mathematica* Player package available at http://www.wolfram.com/products/player/download.cgi. You can repeat the calculations for different events, play the animations, etc. If you have access to a *Mathematica* 6.0 license (most Universities and High Energy Physics Laboratories do) then you can also modify the calculations, choose different assumptions about detector resolutions, etc.

## Acknowledgements:

I would like to thank Brian Cox, Sergio Rescia, Dino Goulianos and Mike Zeller for useful discussions. I would like to thank Steven Wolfram for making scientific computing fun and James Mulnix of Wolfram Research technical support and Sergio Rescia of BNL for helping me up a fairly rapid learning curve. This work has been supported under DOE Grant# DE-AC02-98CH10886.

# Appendix 1 Mathematica calculations:

- **Basic Fomulae for determining the Luminosity function L (z, t) = L (z)*L (t). The beam cross section is given by $\sigma_{x,y}{}^2 = \epsilon_{x,y}\,\beta(z)/(6\pi\gamma)$, where $\epsilon$ is the transverse emittance of the beam, the betatron amplitude function, $\beta(z)$, is the one for a drift since there are no focusing elements in the region of interest and $\gamma$ is the usual relativistic factor. Since the luminosity is $\sim I_1\, I_2 /(\ \sigma_x\, \sigma_y\ )$ , it is weighted by $1/\beta$. We use units of centimeters and nanoseconds throughout.**

```
c = 30;
It[z_, t_] := PDF[NormalDistribution[c * t, 7.55], z]
```

```
betastar = 55;
```

```
30
```

```
Betatron[z_] := betastar + z * z / betastar
```

```
Plot[Betatron[z], {z, -20, 20}]
```

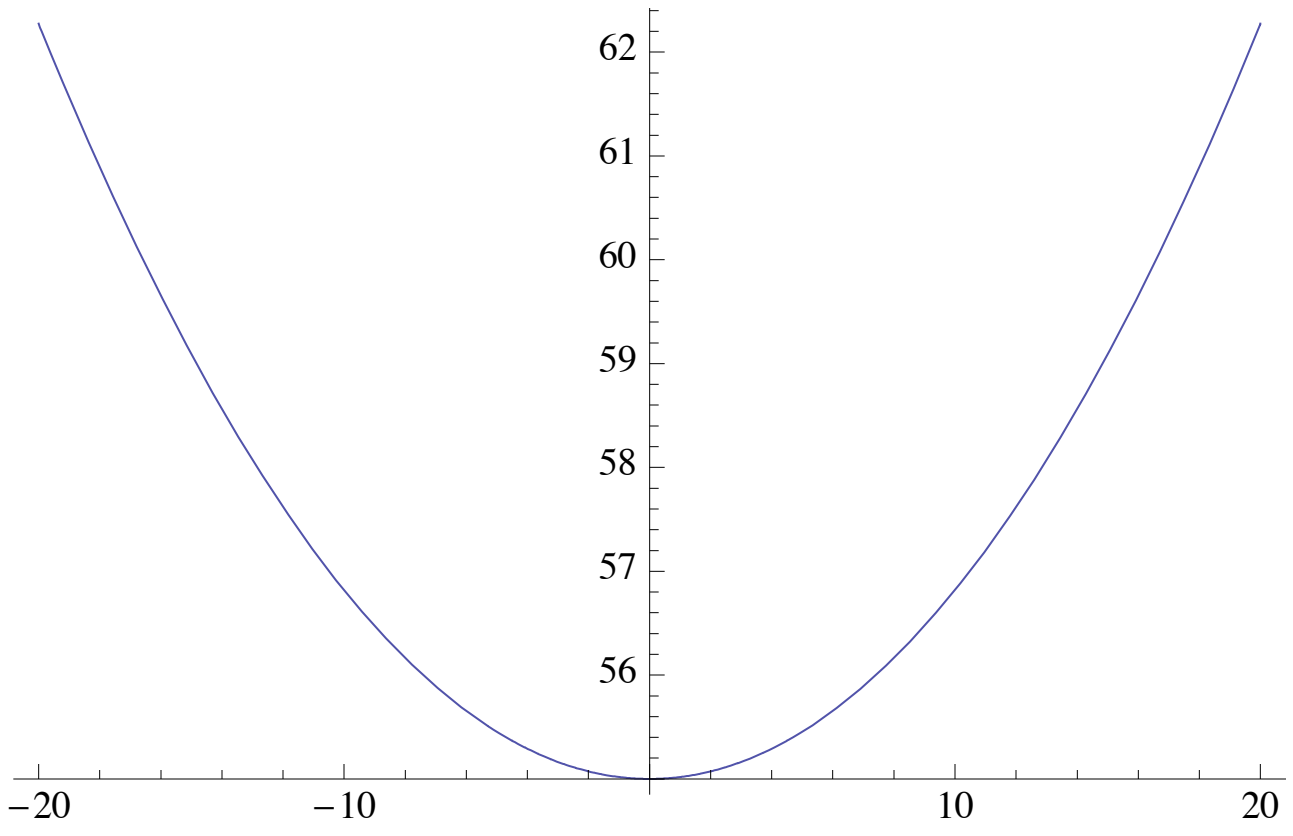

- **A movie of the Luminosity Function for one crossing**

Below, we set `SaveDefinitions->True` so that the movie plays even if previous cells haven't been evaluated.

```
Animate[Plot[{It[z, t] * It[z, -t] / Betatron[z], 0.0008 * PDF[NormalDistribution[0, 4.82], z]},
  {z, -30, 30}, PlotRange → Full], {t, -1, 1}, SaveDefinitions → True]
```

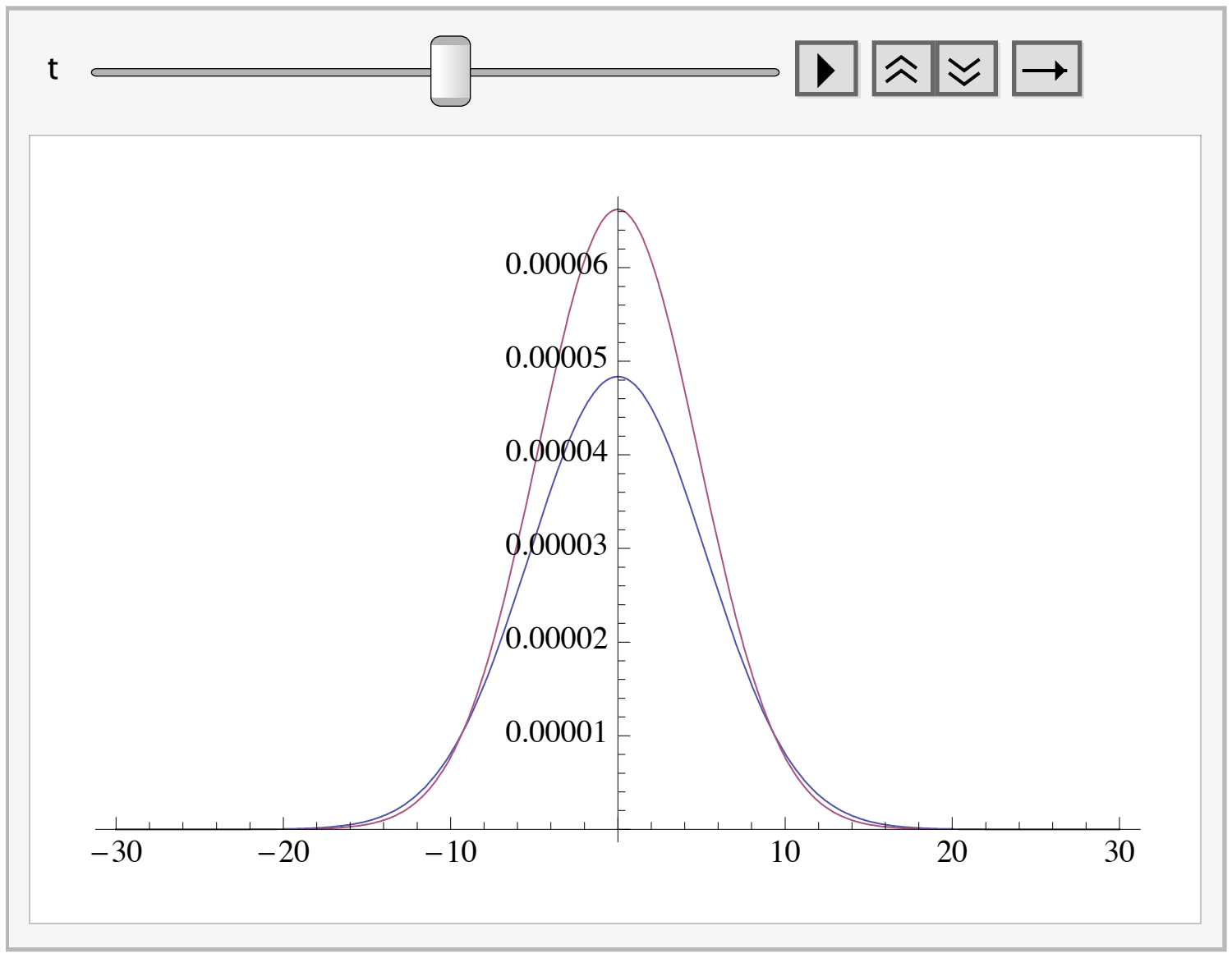

## Now Calculate the proper RMS of the space distribution including Betatron Amplitude Function and the crossing angle (vertical at Atlas and horizontal in CMS) of 285 $\mu$radians

Added a pattern match to definition of **`L`** below. This allows us to remove **`Quiet`** from definition of **`sigmalum`**. (The error messages were caused because the **`NIntegrate`** in the definition of **`sigmalum`** does some pre-processing in which it calls **`L`** with a symbolic parameter, thus causing a complaint from the called **`NIntegrate`**. The pattern match causes **`L`** to call **`NIntegrate`** only when it recieves an actual numerical parameter.)

```
L[z_?NumericQ] := NIntegrate[It[z, t] * It[z, -t] / Betatron[z], {t, -10, 10}]

sigmalum = Sqrt[NIntegrate[z * z * L[z], {z, -20, 20}] / NIntegrate[L[z], {z, -20, 20}]]

5.28402
```

```
sigmay = Sqrt[2] * 16.7 * 0.0001;
theta = 285 * 0.000001;
Clear[Lp];
Lp[z_?NumericQ] := L[z] * NIntegrate[PDF[NormalDistribution[y - theta * z / 2, sigmay], y] *
     PDF[NormalDistribution[y + theta * z / 2, sigmay], y], {y, -0.5, 0.5}];
Plot[Lp[z], {z, -20, 20}, PlotRange → Full, Frame → True,
 FrameLabel -> {"Z in Centimeters", "Luminosity", "Final Luminosity profile in space", ""}]
```

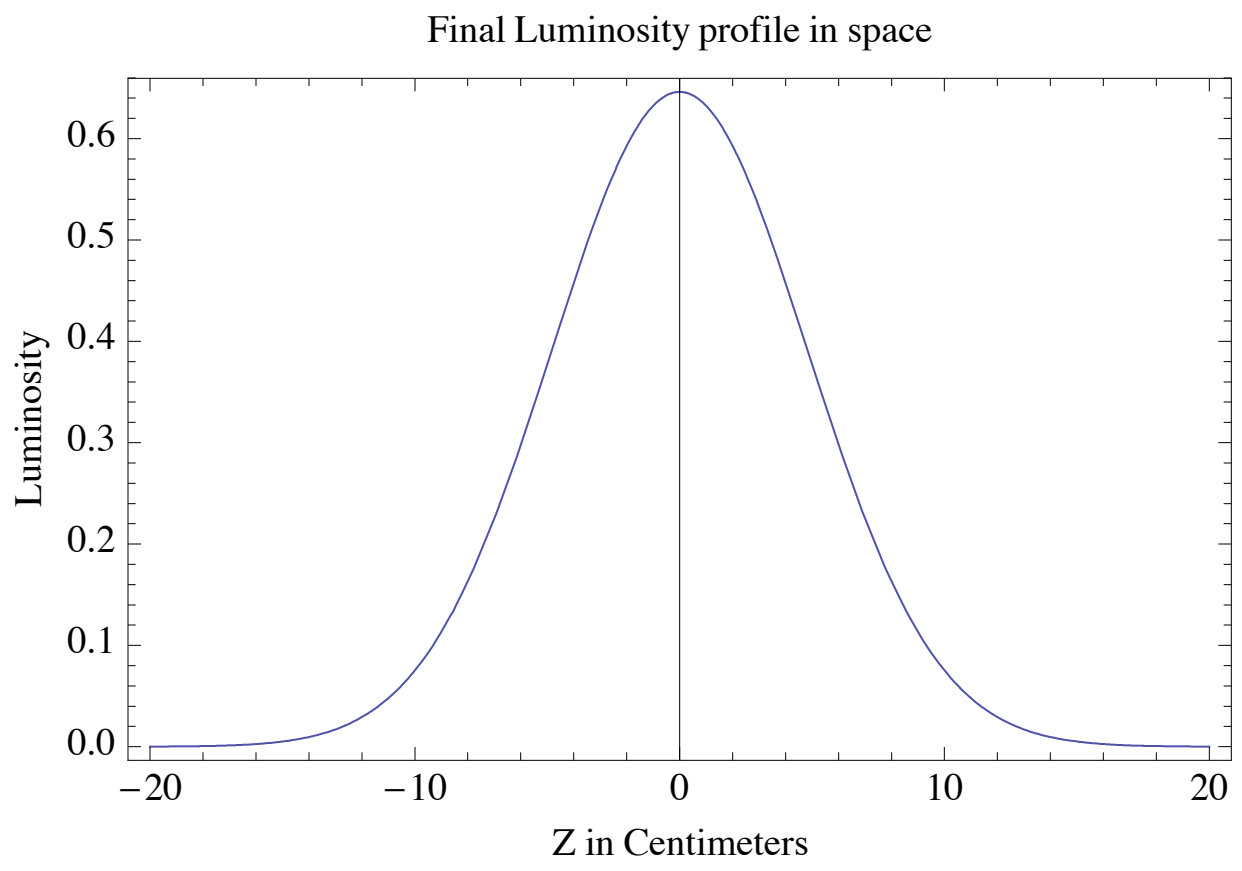

## Fig. 1 Luminosity profile along z integrated over time and including machine optics as well as crossing angle

```
sigmalumi = Sqrt[NIntegrate[z * z * Lp[z], {z, -20, 20}] / NIntegrate[Lp[z], {z, -20, 20}]];

sigmalum = sigmalumi

4.82018
```

## Poisson Distribution for the number of Interactions per crossing

```
ListPlot[Table[{k, PDF[PoissonDistribution[20], k]}, {k, 0, 30}], Filling → Axis]
```

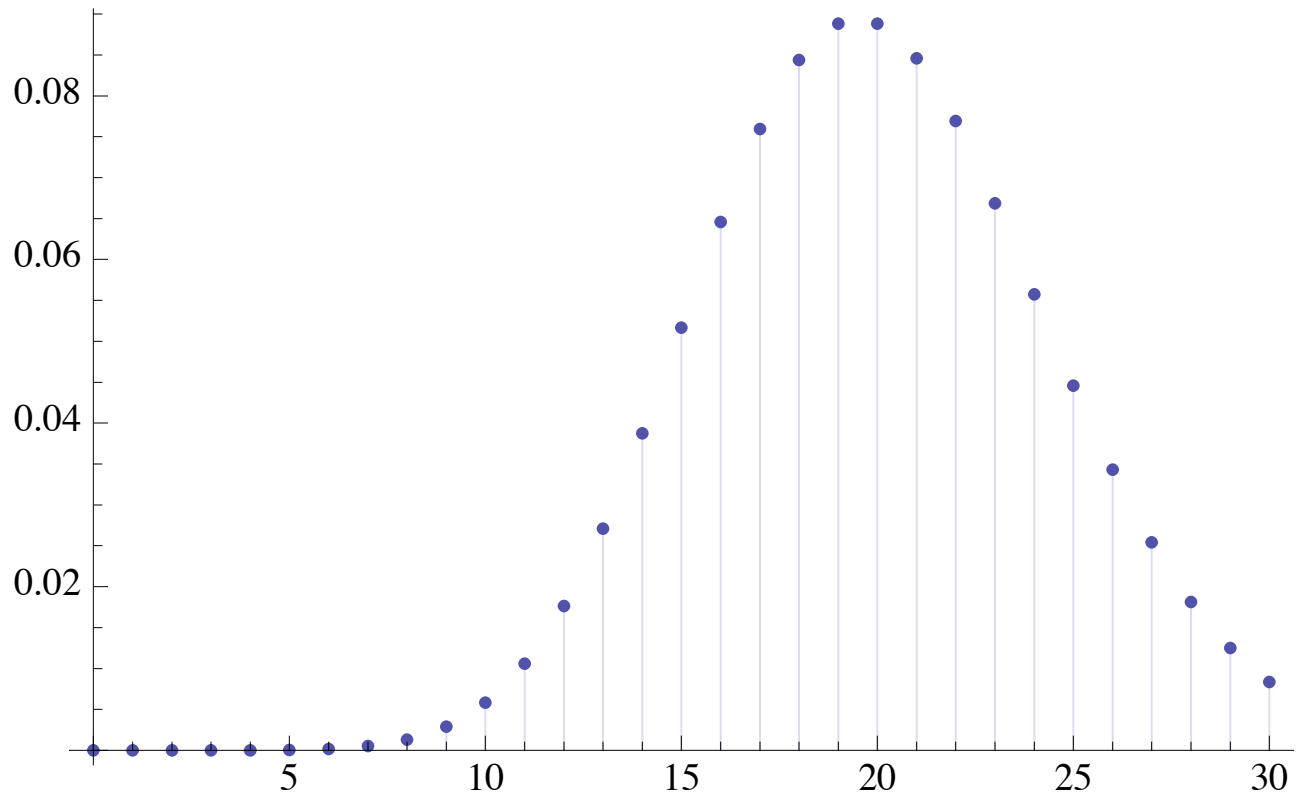

## ■ Now do a Monte Carlo calculation of the Distribution of nearest Neighbor Distances along Z-vertex taking into account the Poisson Distribution on the Number of Events/Crossing

In[66]:=
```
Needs["Histograms`"]
Clear[Tabz];
Tabz = {};
Do[j = RandomInteger[PoissonDistribution[20]];
  Tabnext =
   Differences[Sort[Table[RandomReal[NormalDistribution[0, sigmalum]], {i, 1, j, 1}]]] ;
  Tabmin = Table[Null, {j}];
  Tabmin[[1]] = Tabnext[[1]];
  Do[Tabmin[[k]] = Min[Tabnext[[k]], Tabnext[[k - 1]]], {k, 2, j - 1}] ;
  Tabmin[[j]] = Tabnext[[j - 1]];
  AppendTo[Tabz, Tabmin], {2000}];
Tabz = Flatten[Tabz];
```

In[71]:=
```
Hist = Histogram[Tabz, HistogramRange → {0, 5}, HistogramCategories → 50];
```

In[72]:=
```
Show[Hist, Frame → True, ImageSize → {600, 400}, FrameLabel → {"Centimeters",
   "Frequency", "Distribution of Distances between nearest Neighbors/crossing", ""}]
```

Out[72]=

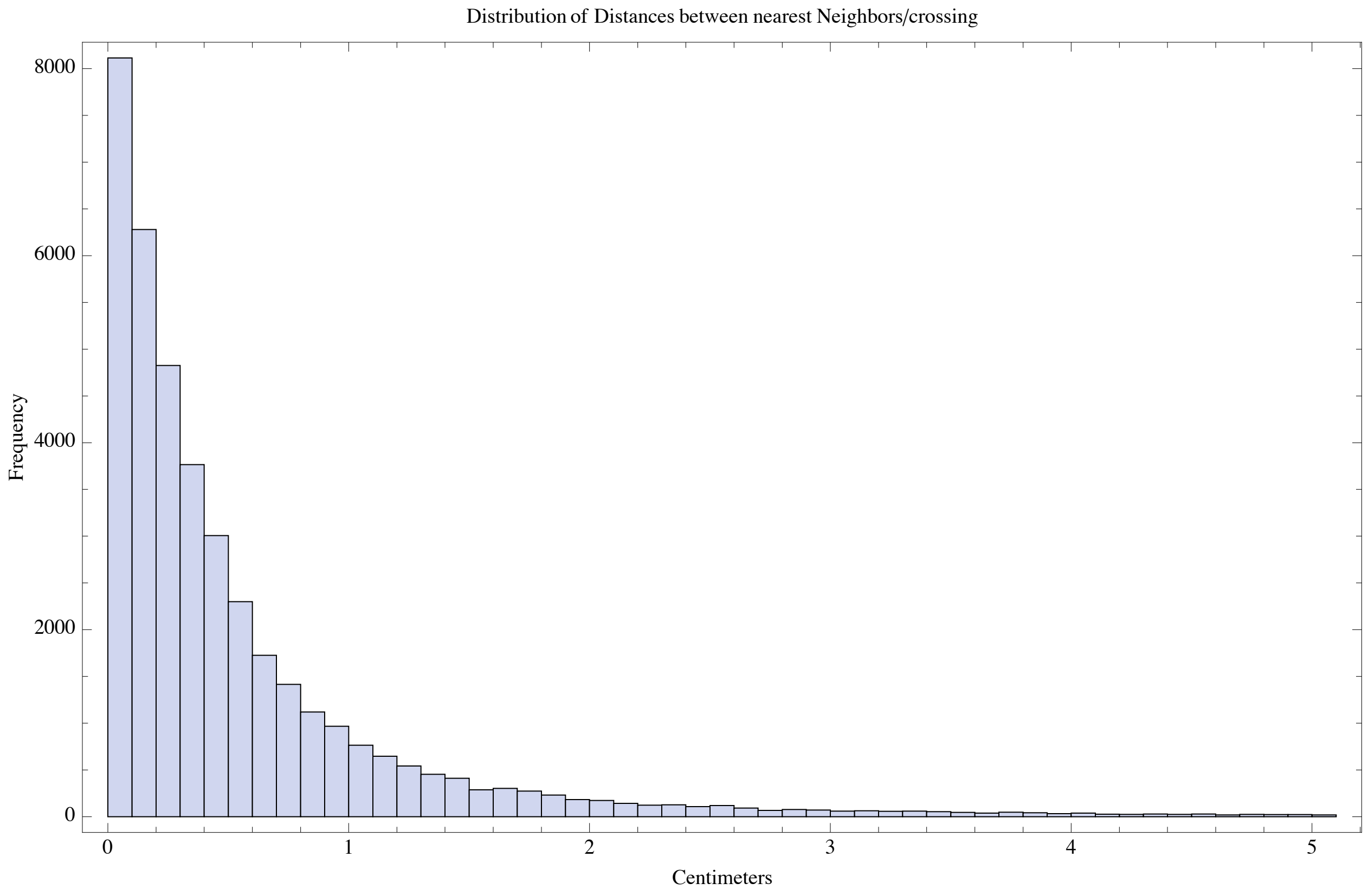

## ■ Fig. 2 Nearest neighbor distances along z

Here we generate a cumulative histogram, using **BinCounts** to sort the data into bins, then **Accumulate** to compute partial sums.

In[73]:=
```
cTabz = Accumulate[BinCounts[Tabz, {0, 5, .1}]];
maxtabz = cTabz[[50]];
```

In[75]:=
```
Histogram[cTabz / maxtabz, ImageSize → {600, 400},
 FrequencyData → True, HistogramRange → {0, 5}, Frame → True,
 FrameLabel → {"centimeters", "Cumulative", "Fig.2 as a cumulative distribution", ""}]
```

Out[75]=

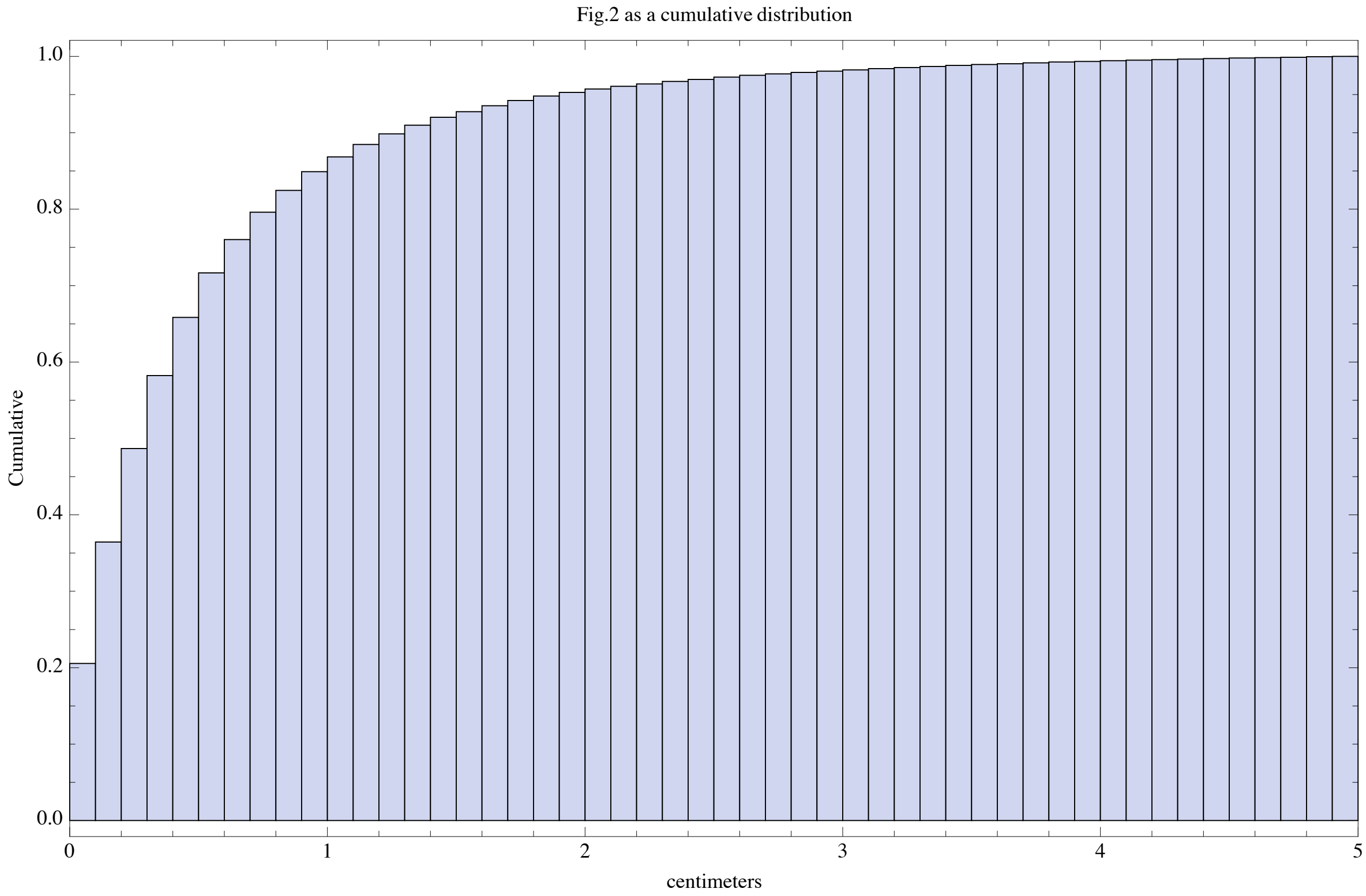

In[76]:=

## ■ Now repeat the same Analysis for the time Domain

Added pattern match to **Ltime**. Removed **Quiet** from **Plot**.

In[100]:=
```
Clear[Ltime];
Ltime[t_?NumericQ] := NIntegrate[It[z, t] * It[z, -t] / Betatron[z], {z, -10, 10}]
```

In[102]:=
```
Plot[Ltime[t], {t, -0.5, 0.5}, Frame → True, PlotRange → Full, ImageSize → {600, 400},
 FrameLabel → {"nanoseconds", "Luminosity", "Luminosity vs. time in the crossing", ""}]
```

Out[102]=

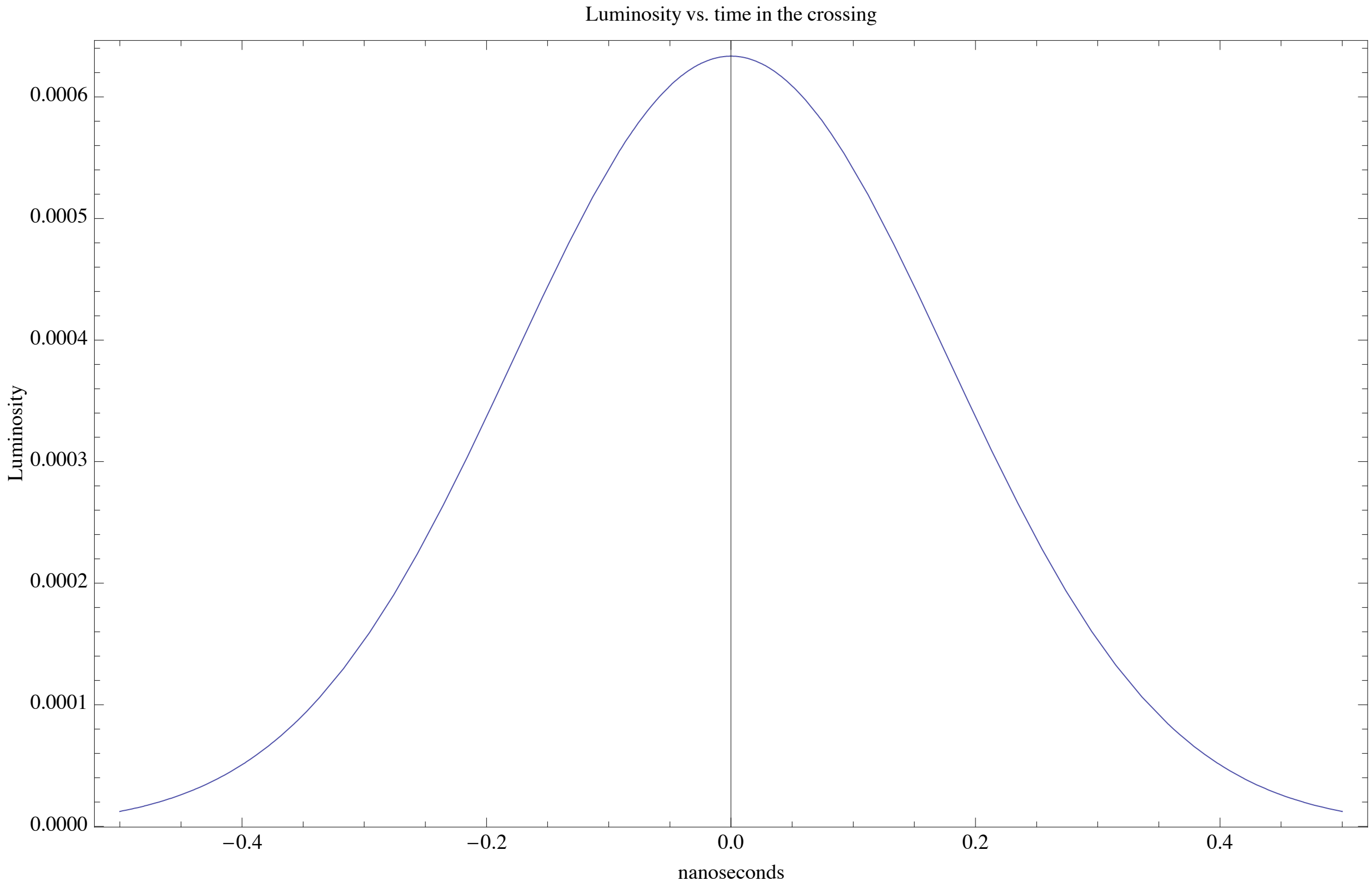

## Fig. 3 Luminosity in time domain

In[123]:=
```
Tabt = {};
sigmatlum = 0.170;
Do[j = RandomInteger[PoissonDistribution[20]];
  Tabnext =
   Differences[Sort[Table[RandomReal[NormalDistribution[0, sigmatlum]], {i, 1, j, 1}]]];
  Tabmin = Table[Null, {j}];
  Tabmin[[1]] = Tabnext[[1]];
  Do[Tabmin[[k]] = Min[Tabnext[[k]], Tabnext[[k - 1]]], {k, 2, j - 1}];
  Tabmin[[j]] = Tabnext[[j - 1]];
  AppendTo[Tabt, Tabmin], {2000}];
```

In[126]:=
```
Tabt = Flatten[Tabt];
Histt = Histogram[Tabt, HistogramRange → {0., 0.2}, HistogramCategories → 20];
```

In[128]:=
```
Show[Histt, Frame → True, ImageSize → {600, 400},
 FrameLabel → {"nanoseconds", "Frequency", "Distance between nearest neighbors/crossing", ""}]
```

Out[128]=

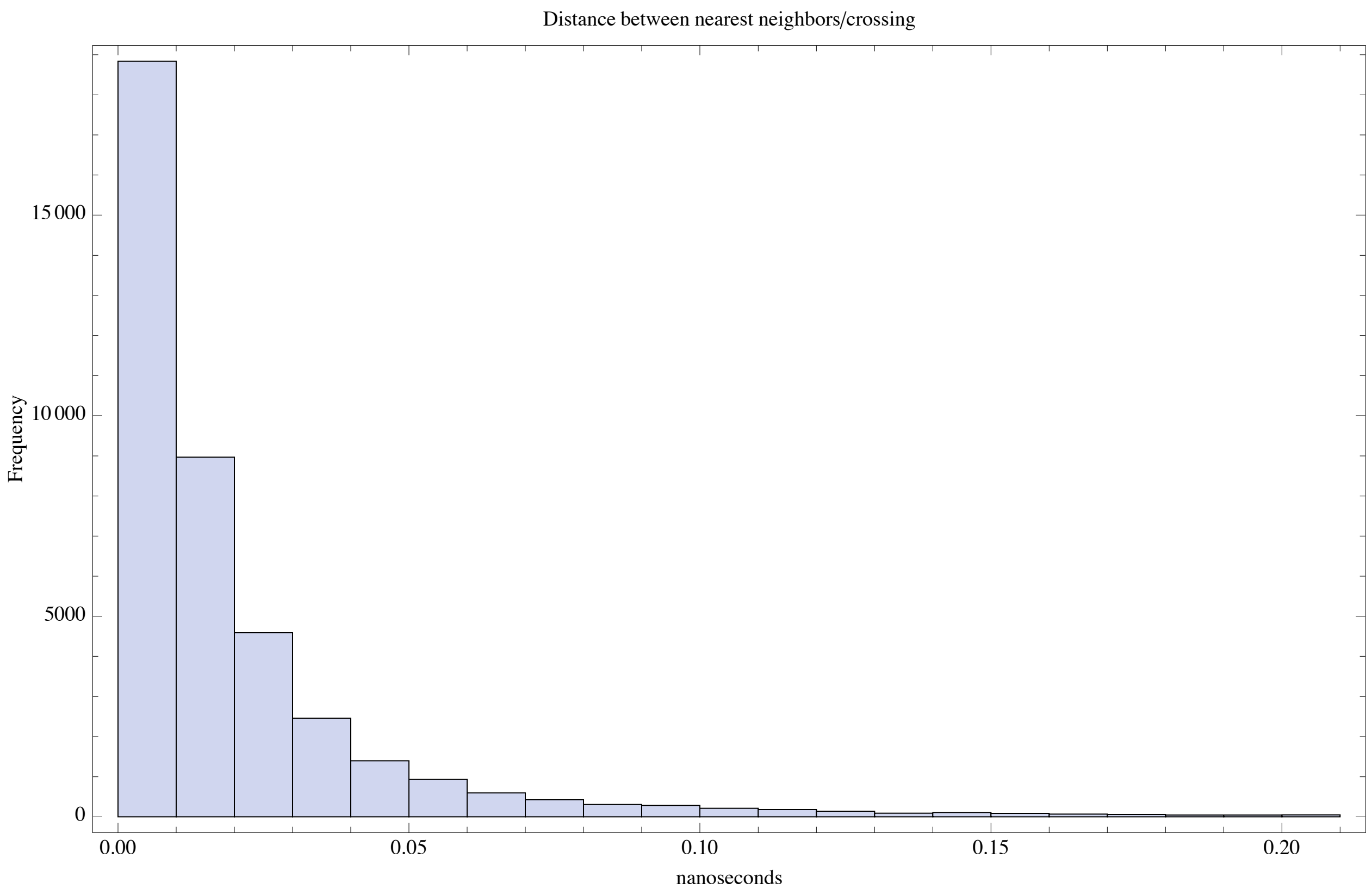

## ■ Figure 4. nearest neighbor distances in time.

In[129]:=

- **Basic Detector Resolutions**

---

- 

---

- **Waveform for the ATLAS Liquid Argon Calorimeter, showing samples used for the Signal processing Algorithm (courtesy of Francesco Lanni).**

---

In[130]:= `Import["/Users/white/Desktop/mathematica_stuff/Signal copy.jpg"]`

Out[130]=

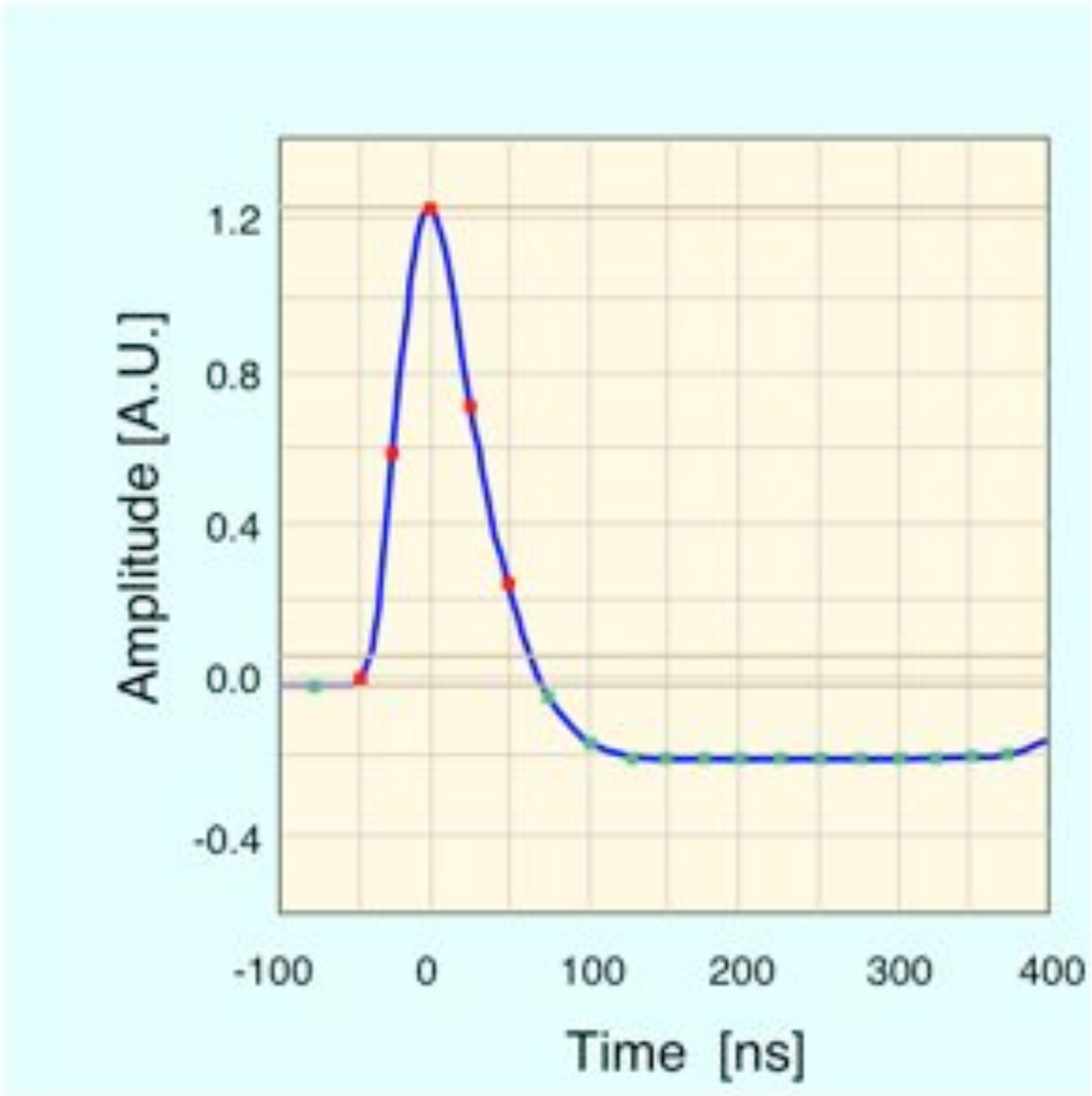

In[131]:=
```
Clear[sigmatLar];
Clear[sigmatof];
sigmatof = {0.01, 0.02};
sigmatLar = {0.01, 0.07, 0.12};
```

In[135]:=
```
Sigmaz[dt_] := dt * c * Sqrt[2] / 2;
Sigmat[dt_] := dt / Sqrt[2];
sigmavertex = Sigmaz[sigmatof[[1]]];
sigmatabspp = Sigmat[sigmatof[[1]]];
SigmatECAL[tECAL_] := tECAL / Sqrt[2];
sigmatoLar = SigmatECAL[sigmatLar[[1]]];
sigmadt = Sqrt[sigmatoLar^2 + sigmatabspp^2];
```

In[142]:=
```
sigma = {sigmalum, sigmatlum};
```

In[143]:=

## Now Generate an Event with 20 Interactions in the Crossing. Points are Events in the Central Detector, Ellipse is the error contour (2 $\sigma$) from the projection of the pp time - of - flight measurement. In this crossing ECAL timing is needed to resolve 2 candidate central events.

```
Coords = {};
Do[AppendTo[Coords,
  Table[RandomReal[NormalDistribution[0, s = Part[sigma, i]]], {i, 1, 2}]], {20}]

TofHit = {};
i = RandomInteger[{1, 20}];
TofHit = Coords[[i]];
TofHit = TofHit + {RandomReal[NormalDistribution[0, sigmavertex]],
    RandomReal[NormalDistribution[0, sigmadt]]};
Plot1 = ListPlot[FindClusters[Coords, 3], PlotStyle → PointSize[0.02]];
Plot2 = Graphics[Circle[TofHit, {2 * sigmavertex, 2 * sigmadt}]];
```

```
Show[Plot1, Plot2, Frame → True, FrameLabel → {"Z-vertex in cms.", "event time(nanoseconds)",
   "One Crossing with 20 Interactions", ""}, ImageSize → {600, 400}]
```

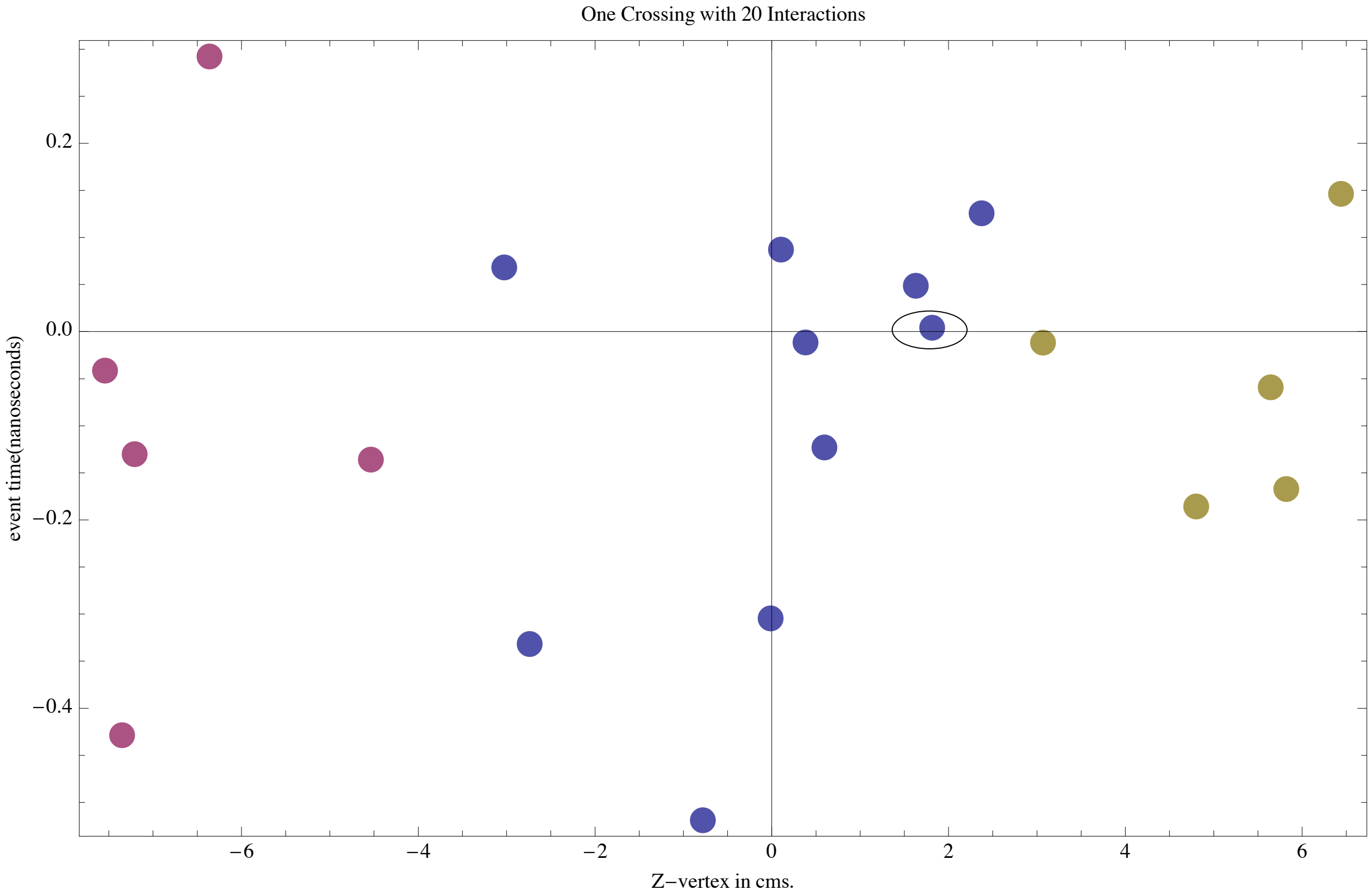

- **Now Repeat this exercise 1000 times per each of 3 assumptions about applied experimental cuts (1, 2, or 3*$\sigma$) for each of 3 cases of ECAL timing**

```
Clear[ideal];
ideal = {};
Clear[good];
good = {};
Clear[modest];
modest = {};
Do[
  sigmadt = Sqrt[SigmatECAL[sigmatLar[[n]]^2 + Sigmat[sigmatof[[1]]]^2]];
  sigmavertex = Sigmaz[sigmatof[[1]]];
  Do[Nevent = 0;
   Do[Coords = {};
    j = RandomInteger[PoissonDistribution[20]];
    Do[AppendTo[Coords,
      Table[RandomReal[NormalDistribution[0, s = Part[sigma, k]]], {k, 1, 2}]], {j}];
    TofHit = {};
    i = RandomInteger[{1, j}];
    TofHit = Coords[[i]];
    TofHit = TofHit + {RandomReal[NormalDistribution[0, sigmavertex]],
       RandomReal[NormalDistribution[0, sigmadt]]};
    If[Position[Coords, {x_, t_} /; (Abs[x - TofHit[[1]]] < m * sigmavertex) &&
         (Abs[t - TofHit[[2]]] < m * sigmadt)] == {{i}}, Nevent = Nevent + 1];, {1000}];
   If[n == 1, ideal = AppendTo[ideal, Nevent / 1000]];
   If[n == 2, good = AppendTo[good, Nevent / 1000]];
   If[n == 3, modest = AppendTo[modest, Nevent / 1000]];
   , {m, 1, 3, 0.5}];
  , {n, 1, 3}];
```

- **Now Plot a Quality Factor which is equal to the Probability that, on average, a unique association is made between an event in the Central Detector with the Leading Protons and that that association is the Correct One.**

```
Plot1 = ListLinePlot[{ideal, good, modest}, ImageSize -> {600, 400},
   DataRange -> {1, 3}, PlotStyle -> {{Thick, Blue}, {Thick, Red}, {Thick, Black}},
   Frame -> True, FrameLabel -> {"experimental cut, sigma",
     "Quality Factor", "Quality Factor as defined in the text", ""}];
Plot2 = Graphics[{Text["σ^t_Ecal=10 psec", {2.5, 0.84}], Text["σ^t_Ecal=70 psec", {2.5, 0.64}],
    Text["σ^t_Ecal=120 psec", {2.5, 0.54}]}];
```

```
Show[{Plot1, Plot2}, ImageSize → {600, 400}]
```

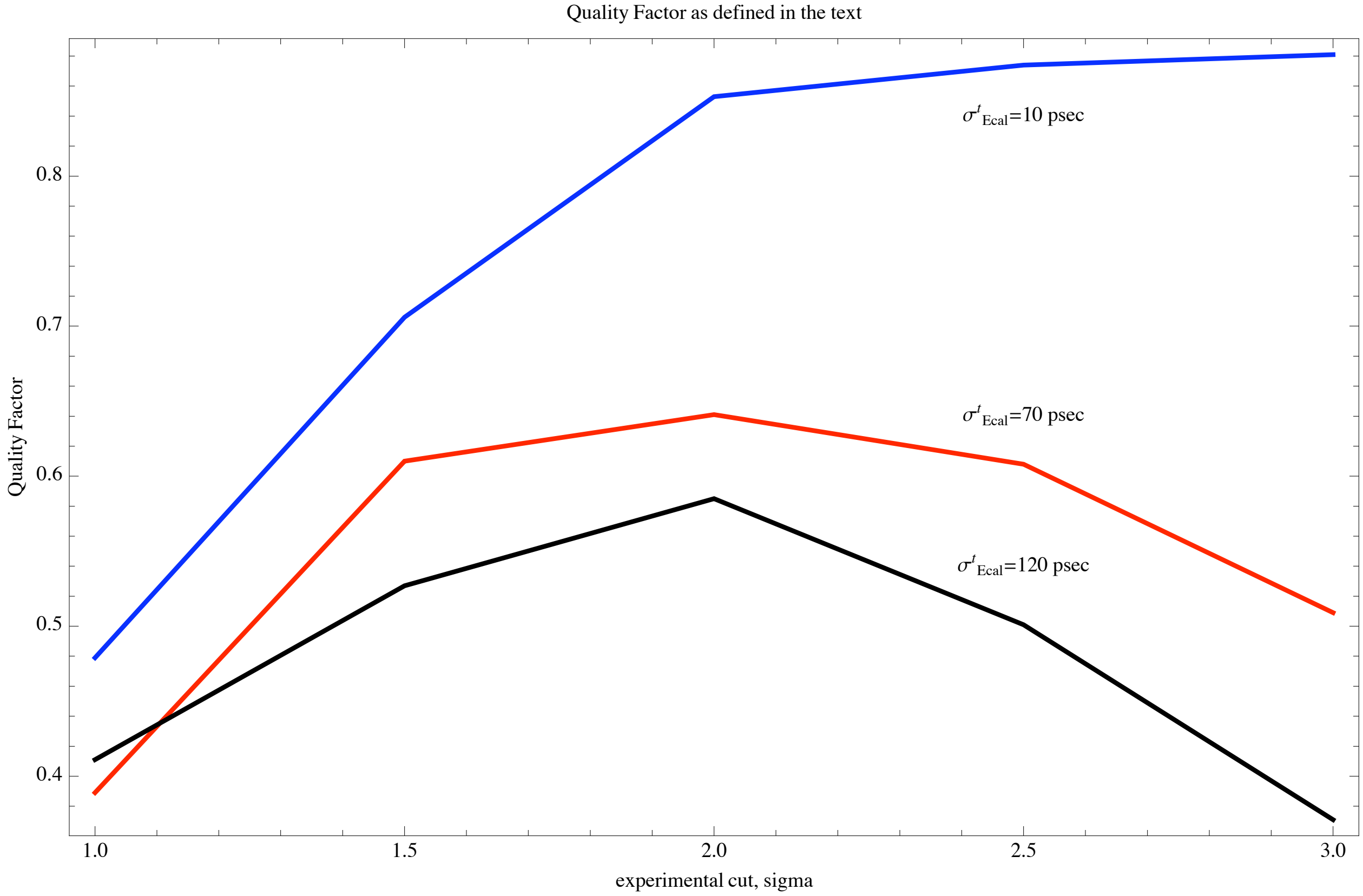

- **Fig. 5 Quality Factor for reconstructing pp->X+pp**

- **Now Plot a Contamination Factor which is equal to the Probability that a random overlapping pair of leading particles from uncorrelated single diffraction dissociation will be associated with a candidate central event with b quark jets in it. We also require that that association is unique.**
  **We do this for three different assumptions about the ECAL timing resolution and plot the result vs. $\sigma$ of the matching cut.**

```
Clear[ideal];
ideal = {};
Clear[good];
good = {};
Clear[modest];
modest = {};
Do[
  sigmadt = Sqrt[SigmatECAL[sigmatLar[[n]]^2 + Sigmat[sigmatof[[1]]]^2]];
  sigmavertex = Sigmaz[sigmatof[[1]]];
  Do[Nevent = 0;
   Do[Coords = {};
    j = 1;
    Do[AppendTo[Coords,
      Table[RandomReal[NormalDistribution[0, s = Part[sigma, k]]], {k, 1, 2}]], {j}];
    Tof1 = RandomReal[NormalDistribution[0, sigmatlum]];
    Tof2 = RandomReal[NormalDistribution[0, sigmatlum]];
    TofHit[[1]] = c * (Tof1 - Tof2) / 2;
    TofHit[[2]] = (Tof1 + Tof2) / 2;
    i = 1;
    If[Position[Coords, {x_, t_} /; (Abs[x - TofHit[[1]]] < m * sigmavertex) &&
          (Abs[t - TofHit[[2]]] < m * sigmadt)] == {{i}}, Nevent = Nevent + 1];, {10 000}];
   If[n == 1, ideal = AppendTo[ideal, Nevent / 10 000]];
   If[n == 2, good = AppendTo[good, Nevent / 10 000]];
   If[n == 3, modest = AppendTo[modest, Nevent / 10 000]];
   , {m, 1, 3, 0.5}];
  , {n, 1, 3}];
```

```
Plot1 = ListLinePlot[{ideal, good, modest}, ImageSize → {600, 400},
   DataRange → {1, 3}, PlotStyle → {{Thick, Blue}, {Thick, Red}, {Thick, Black}},
   Frame → True, FrameLabel → {"experimental cut, sigma",
     "Contamination", "Contamination Factor as defined in the text", ""}];
Plot2 = Graphics[{Text["σ^t_Ecal=10 psec", {2.5, 0.005}], Text["σ^t_Ecal=70 psec", {2.5, 0.03}],
    Text["σ^t_Ecal=120 psec", {2.5, 0.045}]}];
Show[{Plot1, Plot2}, ImageSize → {600, 400}]
```

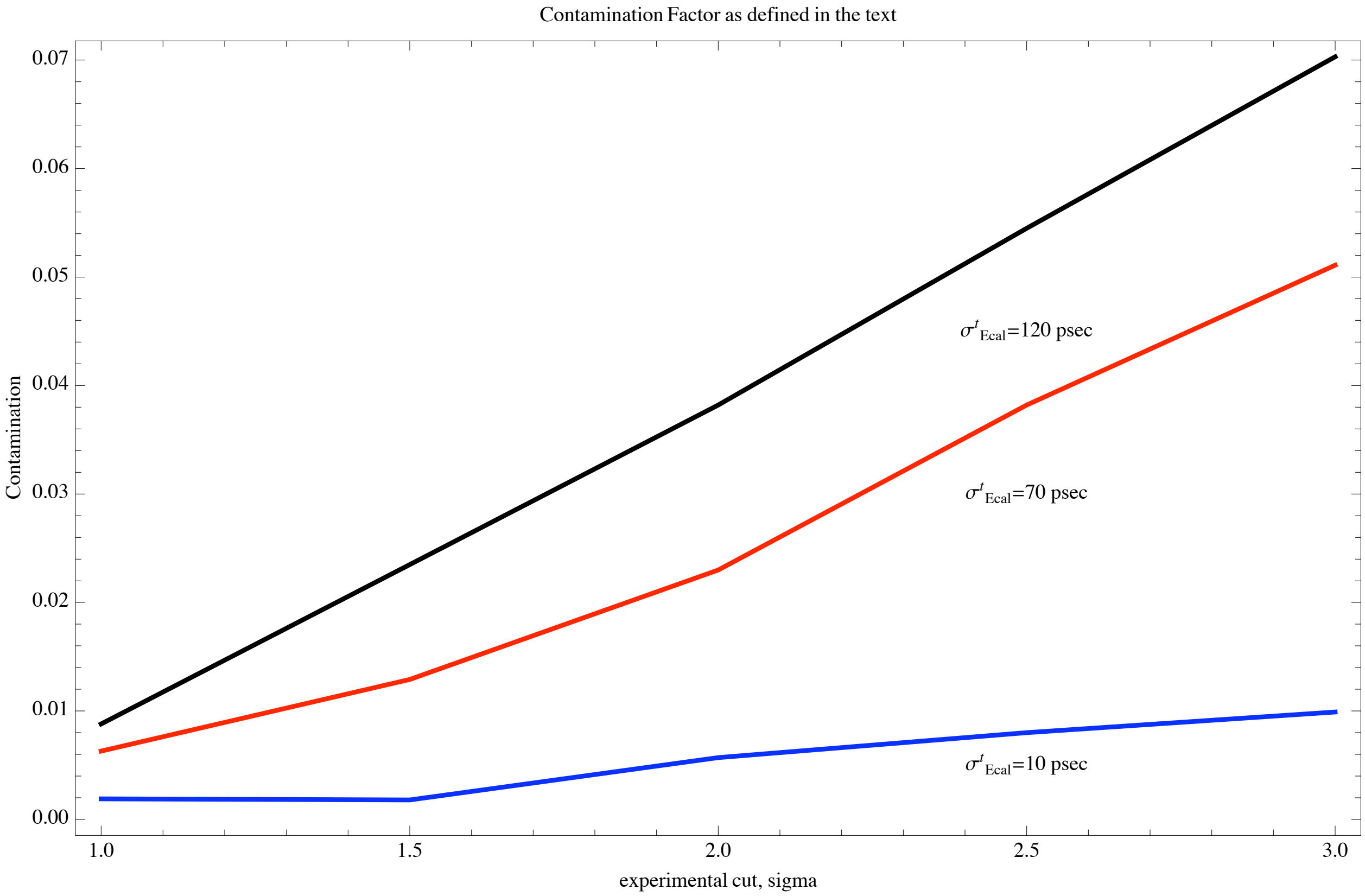

```
{ideal, good, modest}
```

$$\left\{\left\{\frac{1}{1000}, \frac{1}{400}, \frac{41}{10\,000}, \frac{63}{10\,000}, \frac{1}{100}\right\}, \left\{\frac{61}{10\,000}, \frac{137}{10\,000}, \frac{251}{10\,000}, \frac{367}{10\,000}, \frac{527}{10\,000}\right\}, \left\{\frac{103}{10\,000}, \frac{129}{5000}, \frac{101}{2500}, \frac{111}{2000}, \frac{727}{10\,000}\right\}\right\}$$

- **Fig. 6 Probability that a random overlap of two single diffractive leading protons with a Central dijet event will produce an association.**

---

-

- **We now illustrate the importance of a good timing measurement on the leading protons, as opposed to a more modest goal of 20 psec, for example.**

```
Clear[good];
good = {};
Clear[goodbad];
goodbad = {};
Do[
  sigmadt = Sqrt[SigmatECAL[sigmatLar[[2]]^2 + Sigmat[sigmatof[[n]]]^2]];
  sigmavertex = Sigmaz[sigmatof[[n]]];
  Do[Nevent = 0;
   Do[Coords = {};
    j = 1;
    Do[AppendTo[Coords,
      Table[RandomReal[NormalDistribution[0, s = Part[sigma, k]]], {k, 1, 2}]], {j}];
    TofHit = {0, 0};
    tof1 = RandomReal[NormalDistribution[0, sigmatlum]];
    tof2 = RandomReal[NormalDistribution[0, sigmatlum]];
    TofHit[[1]] = c * (tof1 - tof2) / 2;
    TofHit[[2]] = (tof1 + tof2) / 2;
    i = 1;
    If[Position[Coords, {x_, t_} /; (Abs[x - TofHit[[1]]] < m * sigmavertex) &&
         (Abs[t - TofHit[[2]]] < m * sigmadt)] == {{i}}, Nevent = Nevent + 1];, {5000}];
   If[n == 1, good = AppendTo[good, Nevent / 5000]];
   If[n == 2, goodbad = AppendTo[goodbad, Nevent / 5000]];
   , {m, 1, 3, 0.5}];
  , {n, 1, 2}];


Plot1 = ListLinePlot[{good, goodbad}, ImageSize → {600, 400},
   DataRange → {1, 3}, PlotStyle → {{Thick, Blue}, {Thick, Red}},
   Frame → True, FrameLabel → {"experimental cut, sigma", "Contamination",
     "Contamination Factor for 10 and 20 psec tof resolution", ""}];
Plot2 = Graphics[{Text["σ^t_TOF=20 psec", {2.5, 0.06}], Text["σ^t_TOF=10 psec", {2.5, 0.03}]}];
```

```
Show[{Plot1, Plot2}, ImageSize → {600, 400}]
{good, goodbad}
```

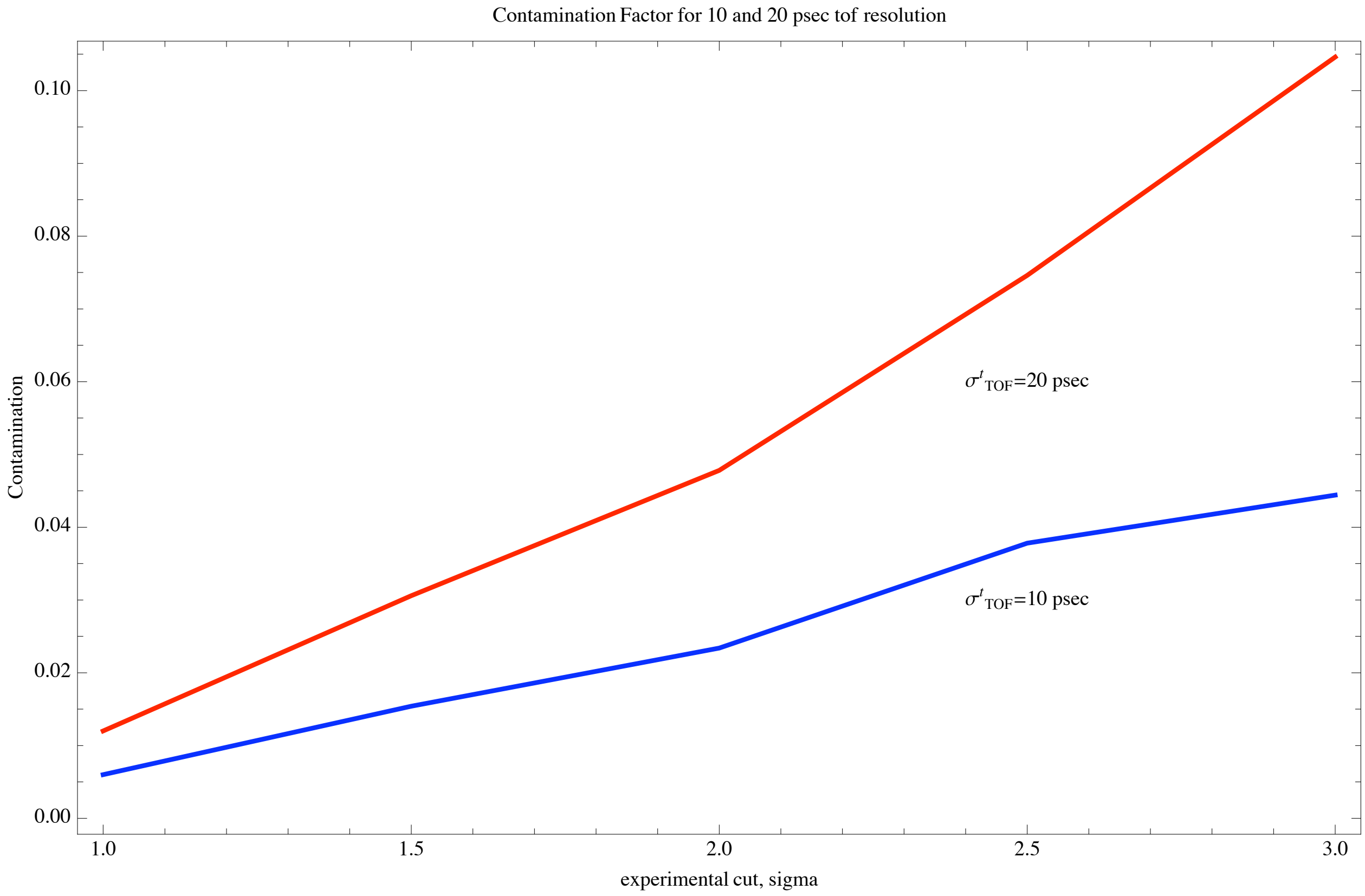

$$\left\{\left\{\frac{3}{500}, \frac{77}{5000}, \frac{117}{5000}, \frac{189}{5000}, \frac{111}{2500}\right\}, \left\{\frac{3}{250}, \frac{153}{5000}, \frac{239}{5000}, \frac{373}{5000}, \frac{523}{5000}\right\}\right\}$$

- **Fig. 7 Probability that a random overlap of two single diffractive leading protons with a Central dijet event will produce an association evaluated for 10 psec prton tof resolution (lower curve) and 20 psec tof resolution (upper curve).**

---

# The End

---